\documentstyle[12pt]{article}
\font\bbigtenbf=cmbx10 scaled \magstep3

\font\twelverm=cmr10 scaled  \magstep2

\def\be{\begin{equation}}
\def\ee{\end{equation}}
\def\bea{\begin{eqnarray}}
\def\eea{\end{eqnarray}}
\def\bea*{\begin{eqnarray*}}
\def\eea*{\end{eqnarray*}}

\def\math{\mathsurround 0pt}
\def\oversim#1#2{\lower.5pt\vbox{\baselineskip0pt \lineskip-.5pt
        \ialign{$\math#1\hfil##\hfil$\crcr#2\crcr{\scriptstyle\sim}\crcr}}}

\def\({\left(} \def\){\right)}
\def\[{\left[} \def\]{\right]}
\def\d{\partial}
\def\half{{\mathchoice{{\textstyle{1\over 2}}}{1\over 2}{1\over 2}{1 \over 2}}}
\def\unit#1{\ifinner \;
            \else \quad \fi
            {\rm #1}}

\def\dbw{\mathop{\d}\limits^{\leftrightarrow}}

\def\bogo{{Bogomol'nyi}}
\def\NO{{Nielsen-Olesen}}
\def\sm{{$\sigma$-model}}
\def\Phibar{\bar\Phi}
\def\F{{\cal F}}

\begin{document}
  \begin{titlepage}
  \begin{flushright}
        {DAMTP-HEP-92-24}\\
        {NSF-ITP-92-75}\\
        {May 1992 }\\
  \end{flushright}

\vskip 1 true in
\begin{center}
{\bbigtenbf Semilocal Topological Defects}\\
\bigskip
\bigskip
{\twelverm Mark Hindmarsh}\\
\medskip
{\it
\baselineskip=15pt
{Department of Applied Maths and Theoretical Physics}\\
{University of Cambridge}\\
{Cambridge CB3 9EW}\\
{U.K.}\\}
\medskip
and\\
\medskip
{\it
Institute for Theoretical Physics\\
University of California\\
Santa Barbara CA 93106\\
U.S.A.\\
}
\end{center}
\vskip 1 true in
\vfill

\begin{abstract}
Semilocal defects are those formed in field theories with spontaneously
broken symmetries, where the vacuum manifold $M$ is fibred by the
action of the gauge group in a non-trivial way.  Studied in this paper
is the simplest such class of theories, in which $M\simeq S^{2N-1}$,
fibred by the action of a local $U(1)$ symmetry.  Despite $M$ having
trivial homotopy groups up to $\pi_{2N-2}$, this theory exhibits a
fascinating variety of defects: vortices, or semilocal strings;
monopoles (on which the strings terminate); and (when $N=2$)
textures, which may be stabilised by their associated magnetic field to
produce a skyrmion.
\end{abstract}

\end{titlepage}

\section{Introduction}
The spontaneous breaking of symmetry lies at the heart of modern
particle physics, for if we believe that nature is more symmetric at
higher energies, then we must also have an explanation for the apparent
lack of symmetry in the universe today.  The success of the partial
unification of the electromagnetic and weak forces in the framework of
the Salam-Weinberg gauge theory \cite{WS} has led to many proposals
for enlarging the symmetry groups of matter and forces, perhaps most
famously Georgi-Glashow SU(5) unification \cite{geo-qui}.  Global
symmetries have also been suggested, notably the axial \cite{pec-qui}
and family \cite{wil} ones.  When we combine these ideas with the
paradigm of the hot Big Bang \cite{BigBang}, finite temperature field
theory tells us that the universe would have gone through a series of
phase transitions \cite{PhaseTranEarlyU} at which more of the
symmetries would have become hidden or broken (although there are
proposals for symmetry increase with reduced temperature
\cite{lan-pi}).  In the past decade or so it has become clear that
the effects of the phase transitions can be felt long after they
finish, through the formation of topological defects
\cite{vil85,bra89} and massless Nambu-Goldstone modes
\cite{tur-spe91}.  In fact, these phenomena are almost the only way
of deriving information about the unification scale of $10^{15}$ GeV.

Topological defects, which may result from the spontaneous breaking of
global or local symmetries, can briefly be characterised as regions of
space-time with dimension $d<4$ in which the order parameter of
the phase transition vanishes due to the topological properties of the
broken and unbroken symmetry groups.  In decreasing order of their
dimension, they are termed domain walls \cite{vil85}, strings or
vortices \cite{vil85,bra89}, monopoles \cite{Mono}, and
finally textures \cite{dav87,Tex} or sphalerons
\cite{Sphal}.  These last are single spacetime points where the order
parameter goes through zero.
Related to this last class are skyrmions
\cite{Skyrm}, which can be thought of as textures prevented from
collapsing by higher derivative terms in the action.  Nambu-Goldstone
modes result from the breaking of global symmetries only, and they seem
to possess an important property which allows them to persist long
after the phase transition is complete:  scaling \cite{tur-spe91}.
Because they are massless, the only relevant length scale for their
dynamics in the early universe is the Hubble length $H^{-1}$.  The
direction of the order parameter $\phi$ is correlated over this
distance, so at any epoch the energy density in the Nambu-Goldstone
modes is $\sim \eta^2H^2$, where $\eta = |\phi|$. The density parameter
for these modes is then roughly $G\eta^2$, where $G$ is the
gravitational constant.  This scaling property is also thought to be
shared by strings (both local and global) \cite{StringScal}, global
monopoles \cite{GlobMono}, and global textures \cite{Tex}.  If this
is the case, then any of these objects will produce a scale-free
spectrum of density perturbations.  It is tempting to regard it as more
than a coincidence that the required amplitude for the formation of
structure in the universe is obtained if $\eta$ is about the Grand
Unification scale.  The energy in the Nambu-Goldstone modes and in
topological defects may also account for the recently observed
fluctuations in the Cosmic Microwave Background \cite{COBE}.  Other observable
effects include gravitational lensing (principally from strings) and
the production of a significant density of radiation, either
gravitational or Nambu-Goldstone.  The radiation density is bounded by
the expansion rate of the universe during nucleosynthesis
\cite{dav85}, in the same way that the number of neutrino species is
bounded (see {\it e.g.} Kolb and Turner \cite{BigBang}).

The above considerations therefore motivate the study of topological
defects, particularly strings, monopoles, and textures, and also global
symmetry breaking, in a cosmological context.  A piece of ideology
picks out the class of models considered in this paper:  it is that
nature should accomplish its phase transitions in the most economical
way, with the minimum of scalar fields.  The scalar sector of most GUTs
so far proposed is made complicated enough in the effort to reduce the
gauge symmetries.  This leads one naturally to the study of models in
which both global and local symmetries are broken at the same phase
transition.

In this paper, the simplest class of models which exhibit this property
is studied.  One might term them extended Abelian Higgs (EAH) models,
for they consist of $N$ scalar fields with a $U(N)$ symmetry, but only
with the overall phase gauged, leaving an $SU(N)$ global symmetry.
When $N=1$ we have the ordinary Abelian Higgs model.  It has already
been shown that this class has some unusual properties, and it contains
some general lessons for the study of phase transitions and the
associated topological defects.  Perhaps the most remarkable point is
that the models exhibit stable vortices for all $N$, despite the first
homotopy group of the vacuum manifold being trivial when $N>1$
\cite{vac-ach,hin}.  This is made all the more interesting by
the observation made by Vachaspati \cite{vac92} that the $N=2$ case
represents the scalar sector of the Electroweak theory in the limit
that the Weinberg angle is $\pi/2$, and that the vortices retain their
stability away from this limit.  The stability of the vortices of the
EAH models was examined in Ref \cite{hin}, and it was shown there
that they are stable only if a certain parameter $\beta$, the ratio
between the scalar self-coupling and the square of its gauge charge, is less
than or equal to 1.  The case $\beta$ = 1 is particularly interesting,
for here the equations of motion can be reduced to a first order form,
the \bogo\  equations \cite{bog}.  In this case it was shown in Ref
\cite{hin} that instead of there being a unique vortex solution (the
\NO\ vortex \cite{NO} for $N=1$), there is a family of them labelled
by a complex $(N-1)$-vector.  Furthermore, it was shown in Ref.
\cite{gib+} that there exists a whole family of $n$-vortex solutions
labelled by $Nn$ complex numbers, the extra $2n$ real parameters being
identified with the positions of the centres of the vortices.

In Section 2 of this paper, the model is introduced and the stability
of the vortices demonstrated by numerical methods.  It was pointed out
in \cite{hin} that these vortices bear a very close relation to the
2-dimensional (2D) lumps in ${C}P^{N-1}$ \sm s \cite{raj}.  In Section
3 the connection is made in more detail, it being shown that the
long-wavelength dynamics of the EAH models are those of the \sm.  When
$\beta > 1$, the next term in an expansion in field gradients is shown
to be a Skyrme term \cite{Skyrm}.

In 3 space dimensions there are further unusual features to the EAH
models.  It was surmised in \cite{vac-ach} that because the vacuum
manifold is simply connected, the vortices could end in what was
described as a ``cloud'' of field gradient energy.  In \cite{hin} it
was asserted that this cloud was in fact a sort of global monopole, but
for reasons of space the field configuration was not written down.  In
Section 4 this defect is remedied, and it will be shown that the
structure resembles a Dirac monopole with a physical string.  The
energy of the configuration is of course infinite, as is that of a
single global monopole \cite{GlobMono} or a infinite length of
string.

There is also another 3D solution of interest in the special case $N=2$
where the vacuum manifold is a 3-sphere:  the texture
\cite{dav87,Tex}.  This is explored in Section 5, where it
will be seen that it possesses an associated magnetic field in a
remarkable configuration best described as a twisted vortex.  In the
$\beta\to\infty$ limit, there is the possibility that the magnetic
field pressure can prevent the final collapse and unwinding of the
scalar field.  The magnetic field produces a fourth derivative term in
the effective \sm\ lagrangian -- the Skyrme term in fact -- and thus
the resulting object can be called a skyrmion \cite{Skyrm}.

The question of the formation of all the above objects at a
cosmological phase transition, and their subsequent evolution, is
tackled in Section 6.  The average magnetic flux through a correlated
area can be computed exactly in the $N=2$ case, and from this it is
shown that, even when they are stable, vortices are rather rare.  They
are also likely to be short, terminating at monopoles.  However, when
$\beta<1$ the attraction of monopoles to antimonopoles is argued to be
greater than the string tension, so there is a mechanism for the
creation of string through the annihilation of a pair of oppositely
charged magnetic poles attached to different segments of string.  It is
then plausible that a cosmological scaling solution can be set up, in
which at any epoch the Hubble volume contains a few collapsing
textures, annihilating pairs of poles, and segments of string.  When
$\beta>1$ the dynamics resemble more that of a \sm, since it is
favourable for the scalar field to remain in its vacuum manifold.  In
either case, a scale free spectrum of density perturbations would be
established, although their statistics would probably be quite
different.  It has already be proposed that, separately, string
\cite{vil85,bra89}, global monopoles \cite{GlobMono}, and
global texture \cite{Tex} could seed the structure that we see
today.  The model here presented can produce all three types of
objects, at relatively little cost in complexity, so the notion that it
may be of importance to cosmology is not far-fetched.

In order to back up this notion, some justification for thinking that
such models have some basis in a realistic particle physics theory.  To
this end, a Grand Unified model possessing the required characteristics
is presented in the concluding Section 7, along with some discussion of
the results obtained in this paper.

\section{Extended Abelian Higgs models and their vortex solutions.} The
model studied in the bulk of this paper is a theory of $N$ complex scalars
$\Phi$ with their overall phase gauged and an $SU(N)$ global symmetry.  The
most general renormalisable lagrangian in 4D consistent with these symmetries
is
\be
{\cal L} = |D_\mu\Phi|^2 -\half\lambda(|\Phi|^2 - \eta^2)^2 -
\frac{1}{4}F_{\mu\nu}F^{\mu\nu}
\label{2.1}
\ee
where $D_\mu = \d-ieA_\mu$ and $|\Phi|^2 = \Phibar\Phi$.  At zero
temperature the field has a vacuum expectation value of magnitude
$\eta$, and the symmetry of the vacuum is reduced to a global
$U(N-1)$.  In particular, the gauge symmetry is broken, and the
physical spectrum consists of a vector particle with mass $m_v =
\sqrt{2}e\eta$, $2(N-1)$ Nambu-Goldstone bosons, and a Higgs
scalar with mass $m_s = \sqrt{2\lambda}\eta$.  The symmetries of the
potential are stable to radiative corrections, because $|\Phi|^2$ is the
only invariant consistent with the $SU(N)\times U(1)$ invariance.
The issue raised in
Refs.  \cite{vac-ach,hin} was whether there are any
non-trivial classical solutions to the equations of motion following
from (\ref{2.1}).  It is clear that there are when $N=1$, for then the
theory reduces to the ordinary Abelian Higgs model, for which it is
known that there are relativistic flux tubes or \NO\ vortices
\cite{NO}.  In fact, any classical solution of the Abelian Higgs
model carries over to the EAH models.  This can immediately be seen
from the equations of motion
\be
\begin{array}{rcc}
D^2\Phi +\lambda(|\Phi|^2 - \eta^2)\Phi &=& 0 \\
\d^\mu F_{\mu\nu} + ie\Phibar{\dbw}_{\nu}\Phi - 2e^2A_\nu|\Phi|^2 &= &0
\end{array}
\label{2.2}
\ee
which are solved by any fields of the form $(A_\mu,\phi\Phi_1/\eta)$,
where $\Phi_1$ is a constant unit vector, if $(A_\mu,\phi)$ are solutions of
the ordinary Abelian Higgs model.  However, it does not necessarily
follow that they are stable solutions, for it may be that a small
perturbation away from the embedded Abelian Higgs model solution
grows.
{}From another point of view, one might not expect there to exist vortex
solutions at all, for the conventional indicator of the presence of
vortices, the first homotopy group of the vacuum manifold $M$, is
trivial for all $N>1$.  However, we shall see that this fact is to be
interpreted with some caution, because (as shown in \cite{hin})
stable vortex solutions, termed ``semilocal strings'' in Ref.
\cite{vac-ach}, do exist, but only when $\beta \equiv \lambda/e^2  <
1$.

The
simplest vortex solution is of course a static straight string, which can be
chosen to lie on the $z$ axis with fields independent of $z$.  Accordingly,
this dimension will be ignored for the moment, and the transverse dimensions
shall be denoted $x^i$ with $i \in \{1,2\}$.  Searching for static solutions is
equivalent to minimising the energy (per unit length) functional
\be
{\cal E} = \int d^2x \[|D_i\Phi|^2 + \half\lambda(|\Phi|^2 - \eta^2)^2\ +
\half B^2\]
\label{2.3}
\ee
where $B=\epsilon_{ij}\d_iA_j$ is the ($z$ component of) magnetic
flux.  If ${\cal E}$ is to be finite, each term in (2.3) must
seperately vanish faster than $r^2$ at infinity.  The vanishing of the
potential energy density implies firstly that $|\Phi|\to\eta$, {\it
i.e.} that $\Phi$ lies in $M$ at $r=\infty$.  The most important
observation from our point of view is that the covariant derivative
term must also vanish at large $r$, which implies that $\Phi(x^i)$ must
be a gauge transformation: that is,
\be
\Phi(x^i) \to \exp\(ie\int_{x_0^i}^{x^i}dy^iA_i(y)\)\Phi(x_0^i)
\label{2.4}
\ee
when $|x_0|$ and $|x|$ tend to infinity.
The continuity of $\Phi$ demands that the integral of the gauge field
around the $z$ axis be an integer multiple of $2\pi/e$.  This
quantisation condition  divides the finite energy field configurations
into topologically inequivalent sectors, but says nothing about the
{\sl existence} of vortex solutions, nor does it guarantee that they
are stable.  The magnetic flux must also disappear as $r\to\infty$,
which imposes a boundary condition on $A_i$, that it must become pure
gauge sufficiently fast.

We will first examine solutions with cylindrical symmetry, for which we may
choose a gauge in which
\be
\Phi \to \Phi_1e^{in_1\theta} \qquad A_i \to n_1\epsilon_{ij}x_j/er^2
\label{2.5}
\ee
For simplicity we shall restrict ourselves to solutions in the $n_1 = 1$
sector, for which the most general solution has the form
\be
\Phi = \sum_{\alpha=1}^{N} f_\alpha(\xi)\Phi_\alpha e^{in_\alpha\theta}
\qquad a_i = \epsilon_{ij}\xi_ja(\xi)/\xi^2
\label{2.6}
\ee
where $\xi^i \equiv e\eta x^i$ are dimensionless coordinates, giving
distances in units of $\sqrt 2$ times the vector particle's Compton
wavelength.  For the rest of this section we shall also rescale the
scalar field $\Phi \to \Phi\eta$ so that it becomes dimensionless.  The
boundary conditions on the fields are
\bea*
f_\alpha \to  \delta_{\alpha 1}, & a \to 1 & {\rm as}\quad  \xi \to \infty \\
   & a \to 0 & {\rm as} \quad \xi \to 0
\eea*
and the $f_\alpha$ vanish at $\xi=0$ if $n_\alpha\ne 0$, which includes of
course $f_1$.  The imposition of cylindrical symmetry means that the
effect of any spatial rotation around the $z$ axis can be undone by a
global $U(N)$ transformation.  This forces the $\Phi_\alpha$ to be
orthogonal, and we may write
\be
\Phibar_\alpha\Phi_\beta = \eta^2\delta_{\alpha\beta}
\ee
As Vachaspati and Ach\'ucarro showed \cite{vac-ach}, there exist
solutions with all the $f_\alpha$ except $f_1$ vanishing, which are
essentially \NO\ vortices with winding number $n_1$.  We denote the
solutions $f_1$ and $a$ in the special case $n_1=1$ by $\bar f$ and
$\bar a$.  \def\fb{\bar f}\def\ab{\bar a} Near the origin $\fb \sim
\xi$ and $\ab \sim \xi^2$, while the behaviour at infinity is
\be
\fb \to 1-c_1\xi^{-\half}\exp(-\sqrt{2\beta}\xi) \qquad
\ab \to c_2\xi^{\half}\exp(-\sqrt{2}\xi)
\label{e:NOasymp}
\ee
with $c_1,c_2$ constants \cite{NO}.  Thus both the scalar and vector fields
tend exponentially to their vacuum values outside regions which are of
order $m_s^{-1}$ and $m_v^{-1}$ in width respectively.

The question raised earlier was whether these solutions are stable or
not.  In order to answer it we must expand the energy functional to
quadratic order around the solution to find the fluctuation operator,
and then look for negative eigenvalues.  Only if there are no negative
eigenvalues is the solution stable.  Generally, this operator is not
diagonal in the perturbations $a_\mu$, $\varphi$, as there can be terms
of the form $\int d^2\xi(\delta^2{\cal E}/\delta
A_i\delta\Phibar)a_i\varphi$ in the expansion.  However, we can
remove these terms by a suitable choice of gauge, one developed by
Moss, Toms and Wright \cite{mos-tom-wri}:
\be
\d_ia_i - i\frac{e}{2} (\Phibar\varphi - \bar\varphi\Phi) + U_ia_i = 0
\ee
where the real vector $U_i$ satisfies the equation
\be
U_i|\Phi|^2 = \half\d_i|\Phi|^2
\ee
In this gauge, the quadratic part of the variation in the energy
functional becomes
\begin{eqnarray}
\delta^2{\cal E} &=& \int d^2\xi \big[\half(\d_ia_j)^2 +
(\d_ia_i)U_ja_j + \half(U_ia_i)^2 + \half a_i^2|\Phi|^2 +\nonumber \\
 & & \qquad |{\bar D}_i\varphi|^2 + \beta(|\Phi|^2 - 1)|\varphi|^2
+ \half\beta(\bar\varphi\Phi+\Phibar\varphi)^2\big]
\end{eqnarray}
where $\bar D$ is the covariant derivative with respect to the
background field $A_i$.  When the background field is an embedding of
the \NO\ vortex, $\Phi^T=(\fb e^{i\theta},0,\ldots,0)$, $A_i =
\epsilon_{ij}\xi_j \ab/\xi^2$, the quadratic variation splits into two
pieces:  one in $a_i$ and the upper component of $\Phi$;
and the other in the remaining components, or
\be
\delta^2{\cal E} = \delta^2{\cal E}_{\rm NO} +
\sum_{a>1}\(|{\bar D}_i\varphi_a|^2 + \beta(\fb^2 -1)|\varphi_a|^2\)
\ee
The subscript NO denotes that this part of the variation is precisely
that obtained from the ordinary \NO\ string, which we know to
be stable against all perturbations $a_i$ and $\varphi_1$.  Therefore
we need consider perturbations in the lower components of the scalar
field only. Because of the global symmetry we may chose the
perturbation in any direction with no loss of generality. Thus writing
$\varphi_a = v_a\sum_n\psi_ne^{in\theta}$ with $v_a$ constant vector,
the question of stability reduces to solving the eigenvalue equations
\be
\[-{1\over\xi}{d\over d\xi}\(\xi{d\over d\xi}\) + {1\over\xi^2}
(\ab-n)^2 + \beta(\fb^2-1)\]\psi_n = \epsilon\psi_n
\label{e:pert}
\ee
If the perturbation operator in square brackets has any negative
eigenvalues, then there will be growing modes, which means that the
\NO\ string is merely a saddle point of the energy functional
${\cal E}$.

Firstly, we establish that there is at most one unstable mode.  Equation
(\ref{e:pert}) has the form of a Schr\"odinger equation with potential
$U_n \equiv (\ab-n)^2/\xi^2 +\beta(\fb^2-1)$.  Now, if $U_n \ge U_m$
for all $\xi$, then the same inequality applies to the lowest
eigenvalues $\epsilon_0(n)$ and $\epsilon_0(m)$.  Therefore if the
lowest eigenvalue with potential $U_n$ is positive, it follows that for
those $m$ satisfying $U_m \ge U_n$ for all $\xi$ there are no negative
eigenvalues.  We know from the stability of the \NO\ vortex that $U_1$
produces no negative eigenvalues, so stable perturbations will be obtained
if
\be
(\ab-n)^2 \ge (\ab-1)^2 \qquad \forall \ab \in [0,1)
\ee
Thus $|n| \ge 1$ give stable perturbations, leaving one possible
unstable mode, which has $n=0$.

I have been able to make further analytic progress only for $\beta =
1$, which shall be described later.  For the rest, therefore, I
resorted to numerical methods.  To solve (\ref{e:pert}) one first must
find the background fields $\fb$ and $\ab$.  To do this I used a simple
relaxation method \cite{Relax}.  The energy functional is discretised
onto a 1D lattice (taken to be 800 points), with derivatives evaluated
on the links, and then minimised at each point $r$ in turn.  This means
solving a cubic equation for $\fb_r$ and a linear one for $a_r$ to
obtain their updated
values.  One can also incorporate an overshoot factor to speed up the
convergence.  For example, if $\tilde{f}_n$ is the solution to $\d{\cal
E}/\d f_n = 0$, then $\fb_n$ is updated to $\fb'_n =
\fb_n+\alpha(\tilde{f}_n - \fb_n)$, with $\alpha>1$.  The energy
converges reasonably fast: typically, after a few hundred iterations it
is changing by a factor of less than $10^{-7}$.  The accuracy of the
algorithm can be checked by calculating $E/2\pi\eta^2$ for the
Nielsen-Olesen vortex with $\beta=1$, for it should be exactly 1 (as
will be explained below).  Using FORTRAN double precision with a step
length $\Delta\xi$ of $1.25 \times 10^{-2}$ the program gives a value
0.999993.  The resulting $\fb$ and $\ab$ were then fed into
(\ref{e:pert}) which was then solved numerically by a shooting method.
The differential equation (11) was discretised by a simple symmetric
scheme whereby $\psi'_r = (\psi_{r+1}- \psi_{r-1})/2\Delta\xi$ and
$\psi'' = (\psi_{r+1}+\psi_{r-1}- 2\psi_r)/\Delta\xi^2$.  Negative
eigenvalues are found by shooting for $\psi_r= 0$ at the boundary
$r=r_{max}$, with initial conditions $\psi_0 = 1$ and $\psi_1 = 1$.  A
Newton-Raphson iteration on the function $\psi_{r_{max}}(\epsilon)$ was
used to automate the convergence.  One can find the lowest eigenvalue
by a judicious choice of the first two values of $\epsilon$, and
checking that the eigenfunction has no nodes.  The results for $\beta$
= 1, 3, 10, 30, and 100 are listed in Table 1.  For $\beta$ = 0.3, 0.1,
0.03, and 0.01 no negative eigenvalues were found.  The accuracy of the
eigenvalues is good:  they have been quoted to 3 decimal places,
because doubling the step length and halving the number of points
changed the lowest eigenvalue for $\beta=10$ from -9.1654 to -9.1653.
The accuracy is not quite as good as this suggests, because as we shall
see next the lowest eigenvalue for $\beta=1$ is not about $4\times
10^{-3}$ but is exactly zero.

In order to prove the existence of this zero mode we need a bit of
machinery, established by \bogo.  He showed that in the special case
$\beta=1$ the second order field equations split up into first order
ones.  This emerges from rewriting the energy functional \ref{2.3}:
\begin{eqnarray}
{\cal E} & = & 2\pi n_1\eta^2+\eta^2\int d^2\xi\[\half|D_i\Phi
+ i\epsilon_{ij}D_j\Phi|^2 + \half(B+|\Phi|^2-1)^2 \right. \nonumber \\
 & & \qquad +\left. \half(\beta-1)(|\Phi|^2-1)^2\]
\end{eqnarray}
where $n_1$ has been taken to be positive, and all currents are assumed
to vanish at infinity.  At $\beta=1$ it can be seen that the energy is
minimised at $2\pi n_1\eta^2$ when the following first order equations
are satisfied:
\begin{eqnarray}
(D_i + i\epsilon_{ij}D_j\Phi)\Phi & = & 0 \\
B + |\Phi|^2 -1 & = & 0
\label{e:bogo}
\end{eqnarray}
In the cylindrically symmetric case with $n_1 = 1$ the \bogo\ equations are
\be
f'_\alpha +{1\over\xi}(a-n_\alpha)f_\alpha = 0 \qquad a' +
\xi(\sum_\alpha f^2_\alpha - 1) = 0
\ee
{}From these equations it is easy to check that the zero mode of the vortex is
\be
\psi = {\cal N} \exp\[-\int_0^\xi{d\xi'\over\xi'}\ab(\xi')\]
\ee
where ${\cal N}$ is a normalisation factor.  This zero mode is a symptom
of an unexpected degeneracy in the solutions to the \bogo\ equations.

Without loss of generality we may take the zero mode to be in the
$\Phi_2$ direction.  Note that the 3 \bogo\ equations are not
independent, for we may replace $f_2$ by $f_1w/\xi$ with any complex
$w$ and still satisfy the equations of motion, providing $f_1$ and $a$
solve
\be
f'_1+{(a-1)\over\xi}f_1=0 \qquad a'+\xi[f_1^2(1+|w|^2/\xi^2)-1] = 0
\ee
If $w=0$ we are left with the \bogo\ equations for the plain Abelian
Higgs model, for which we know there exists a vortex solution.  In Ref.
\cite{hin} it was proved that vortex solutions to (\ref{e:bogo})
exist for {\em any} complex $w$, revealing the \NO\ vortex as just one
of a family of vortex solutions.  Recall now that the choice of
$\Phi_2$ is arbitrary so long as it is orthogonal to $\Phi_1$ and has
unit modulus.  This reveals that the family is labelled by a complex
$(N-1)$-vector ${\bf w} = w\Phi_2$.  The asymptotics of this family for
$w\ne0$ are very different from those of the \NO\ string, whose fields
approach their vacuum values exponentially (\ref{e:NOasymp}).
Expanding $f_1$, $f_2$ and $a$ up to fourth degree in
$\zeta\equiv\xi/w$ it is easily found that
\begin{eqnarray}
f_1 & \simeq & 1 - \half\zeta^{-2} +({\textstyle{3\over 8}}-|w|^{-2})\zeta^{-4}
\nonumber\\
f_2 & \simeq & \zeta^{-1} - \half\zeta^{-3} ,\label{e:SLasymp}\\
a & \simeq & 1 - \zeta^{-2} + (1-4|w|^{-2})\zeta^{-4} \nonumber
\end{eqnarray}
while the magnetic field $B$ tends to
$2|w|^{-2}\zeta^{-4}\equiv2|w|^2\xi^{-4}$.  This power law behaviour is
quite strange: the width of the flux tube is an arbitrary parameter
instead of the Compton wavelength of the vector boson.  This is not at
all what we expect for magnetic fields in the Higgs phase of a theory.

It is clear from (\ref{e:SLasymp}) that $|w|$ controls the width of the
vortex.  The magnitude of the scalar field at the origin also depends
on this quantity, and it is possible to derive the bounds \cite{hin}
\be
1-2|w|^{-2} < f_2(0) < 1
\ee
showing that as the width of the vortex tends to infinity, the
magnitude of $\Phi$ gets closer to 1 everywhere, even at the vortex
core.

The existence of general solutions to the \bogo\ equations
(\ref{e:bogo}) has been proved in Ref. \cite{gib+}, where it was
shown that in the $n$-vortex sector they are labelled by $nN$ complex
numbers.  These numbers can be interpreted as $(N-1)$ internal complex
parameters for each vortex, plus another $n$ labelling their
positions.  There are also time dependent solutions, corresponding to
charged vortices, with their own Bogomol'nyi equations \cite{abr}.

Clearly, the \bogo\ limit $\beta=1$ is rather special, so we are
entitled to ask what happens to the vortices away from this value.  Let
us consider an isolated vortex whose field configuration satisfies the
\bogo\ equations.  Its energy functional can be rewritten
\be
{\cal E} = 2\pi\eta^2 + \half(\beta-1)\eta^2\int d^2\xi B^2
\label{e:nonbogo}
\ee
using the second of equations (\ref{e:bogo}). Recall that at large
distances, $B\to 2|w|^2/\xi^4$:  hence the second term in
(\ref{e:nonbogo}) can be expressed as a function of $w$, or
$\half(\beta-1)\eta^2\kappa(|w|)|w|^{-2}$.  For large $|w|$, $\kappa$
tends to a constant, because the dependence on $|w|$ is merely an
expression of how the magnetic field energy changes under a scale
transformation.  Thus, according to the sign of $(\beta-1)$ the energy
will change with the size of the vortex.  The tendency is to collapse
if $\beta<1$ and expand if $\beta>1$, which is consistent with the
perturbative stability at $w=0$.  In the strict \bogo\ limit it is
possible to approximate the low energy dynamics of the field theory by
slow motion on the space of parameters, or moduli, of the non-trivial
classical solutions \cite{ModSpace}.  For the single vortex, omitting
the trivial centre-of-mass motion, the moduli are the real and
imaginary parts of $w$.  The time dependence of the fields occurs
entirely through the moduli changing with time, so that the infinite
dimensional lagrangian for the fields
\be
L = \int d\xi\[|\dot\Phi|^2\eta^2 + \half(\dot{A}_i)^2\] -{\cal E}
\ee
reduces to a finite dimensional one in $w$.  In the one vortex sector
it is logarithmically divergent, since $\dot\Phi$ contains a piece
$\dot{\bf w}f_1/\xi$.  We can cut off this divergence at a scale $R$,
which might be interpreted as a vortex seperation, to find
\be
L = \eta^2\[T(w)|\dot{\bf w}|^2 - \half\kappa(\beta-1)|{\bf w}|^{-2}\]
\ee
where $T(w)$ includes at piece proportional to $\ln(R/|w|)$.  From this
we can compute the timescale for the collapse or expansion of the
vortex outside the \bogo\ limit.  If we just consider the radial motion
in the moduli space, and ignore the time dependence of the logarithm,
the solutions to the equations of motion are
\be
|w|^2 = |w(t_0)|^2 + \alpha(t-t_0)^2|w(t_0)|^{-2}
\ee
where $\alpha = \kappa(\beta-1)/2\ln(R/|w|)$.  Thus the timescale for
changing size by a factor 2 is approximately $w(t_0)^2/\sqrt\alpha$.
We have been working in units where times and distances are
measured in units of $m_v^{-1}$, so if $W$ is the width in natural units,
the collapse timescale goes as $W^2m_v$.  Thus vortices which are large
compared with the gauge boson Compton wavelength collapse very slowly,
in much more than the light crossing time.

\section{The $CP^{N-1}$ \sm\  as the low momentum limit}
In this section it is shown how the low momentum dynamics of the EAH
models with $N$ scalars approach those of the $CP^{N-1}$
\sm\ \cite{raj}, where by low momentum we mean momenta whose
magnitudes are much less than the masses $m_s$ and $m_v$.  Let us first
describe the finishing point:  \sm s with target space $CP^m$.  These
manifolds are the set of all complex lines in $C^{m+1}$: thus $Z'_a$
and $Z_a$ in $C^{m+1}$ are the same point in $CP^m$ if and only if
there exists a complex number $\alpha$ such that $Z'_a = \alpha Z_a$.
When we use homogeneous coordinates $Z_a$ on $C^{m+1}$ the natural
induced metric on the imbedded $CP^{m}$ is the Fubini-Study metric \cite{law}
\be
g_{ab}(Z) = {\delta_{ab}|Z|^2 - Z_a\bar{Z}_b\over |Z|^4}
\ee
This metric is used to construct the \sm\ action from $m+1$ complex
scalar fields $\Phi_a$:
\be
S_{CP^m} = \int d^4x \(\d_\mu \Phibar_a g_{ab}(\Phi)\d^\mu\Phi_b\)
\label{e:sigmodac}
\ee
It is easily found that this action is invariant under multiplication
of the fields by a space-time dependent complex number, and so it has
both a local scale invariance in the target space and a local
$U(1)$ gauge invariance.  To arrive at equation (\ref{e:sigmodac}) we
first rewrite the gauge potential
\be
A_\mu = -{i\over 2e}{\Phibar{\dbw}_{\mu}\Phi \over |\Phi|^2} + a_\mu
\label{e:A}
\ee
The expression for the gauge field arising from this field redefinition is
\be
F_{\mu\nu} = - {i\over 2e}\[\d_{(\mu}\Phibar\d_{\nu)}\Phi|\Phi|^2 -
(\d_{(\mu}\Phibar\Phi)(\Phibar\d_{\nu)}\Phi)\]/|\Phi|^4 + f_{\mu\nu}
\label{e:F}
\ee
where $f_{\mu\nu} = \d_{(\mu}a_{\nu)}$.  Complex geometers will
recognise the first term on the right hand side as $1/e$ multiplied by
the components of the K\"ahler form $\omega_{\mu\nu}$ in the immersion
$\Phi:R^n \to C^{m+1}$ \cite{law}.  After some algebra the lagrangian
of the EAH model is reexpressed as
\begin{eqnarray}
{\cal L} &=& |\Phi|^2 \d_\mu\Phibar_a g_{ab}\d^\mu\Phi_b -
{1\over 4e^2}\omega_{\mu\nu}\omega^{\mu\nu} +
2(\d_\mu|\Phi|)^2 + e^2a_\mu^2|\Phi|^2 \nonumber \\
 & & \null-{1\over 2e}
\omega_{\mu\nu}f^{\mu\nu} - {1\over 4}f_{\mu\nu}f^{\mu\nu} -
\half\lambda(|\Phi|^2-\eta^2)^2
\label{e:eahlag}
\end{eqnarray}
The degrees of freedom are now explicitly separated.  There is a real
scalar field $|\Phi|$, the Higgs mode, which because of the potential
term tries to keep as close to $\eta$ in magnitude as possible.  This
gives a mass $\sqrt{2}e\eta$ to the propagating modes of the gauge
field $a_\mu$, which in turn react back on $|\Phi|$, with the effect of
reducing its size.  The first term in (\ref{e:eahlag}) can be
immediately identified as the $CP^m$ \sm\ model lagrangian, describing
the NG bosons, provided $\Phi$ remains close to $\eta$.  Again,
spacelike excitations in the $\Phi$ field tend to reduce $|\Phi|$, but
provided the typical momentum ${p}$ in these excitations has $|{p}|^2
\ll \lambda\eta^2 \simeq m_s^2$, they will have little effect.  The
back reaction from the gauge field is slightly more involved.  Consider
the equation of motion for $a_\mu$:
\be
\d_\mu f^{\mu\nu} + 2e^2|\Phi|^2a^\nu = -{1\over e}\d_\mu\omega^{\mu\nu}
\ee
The right hand side acts as a source for $a^\nu$, and is third order in
derivatives in $\Phi$.  Thus in a momentum expansion we can say $a^\nu
= \eta O( p^3/(e\eta)^3)$.  The back reaction of the gauge field on the
Higgs mode can therefore be neglected if $(e\eta)^2(p^3/(e\eta)^3)^2
\ll \lambda\eta^2$, or $p^6/m_v^6 \ll \beta$.

A term not appearing in the usual \sm\ action is that arising from the
gauge field, $-\omega_{\mu\nu}\omega^{\mu\nu}/4e^2 = O(p^4/e^2)$.  It
is consistent to retain this as the next highest order term if it is
larger than $(|\Phi|^2-\eta^2)(\d_\mu\Phibar_ag_{ab}\d^\mu\Phi_b)$.
The equation of motion for the Higgs mode tells us that deviations of
$|\Phi|^2$ from $\eta^2$ are of order $p^2/\lambda$.  Therefore the
condition to be met is $O(p^4/e^2) \ll O(p^4/\lambda)$, or $\beta \ll
1$.  A major result from this section has now been arrived at:  the low
momentum effective lagrangian for the EAH model in the \sm\ limit
($\beta \ll 1$) is
\be
{\cal L}_{\rm eff} = \eta^2 \d_\mu\Phibar_ag_{ab}\d^\mu\Phi_b -
{1\over 4e^2} \omega_{\mu\nu}\omega^{\mu\nu}
\label{e:efflag}
\ee
Recall that the K\"ahler form $\omega_{\mu\nu}$ is the first term on
the right hand side of (\ref{e:F}) multiplied by $2e$.  There is a
special significance to the additional term in (\ref{e:efflag}): it is
effectively a Skyrme term \cite{Skyrm} for the $CP^m$ \sm.  This is
because it is fourth order in all derivatives but only second order in
derivatives with respect to time.  There is yet further geometrical
significance to the relationship between the gauge field and the
K\"ahler form $\omega$.  Firstly, $\omega^m/m!$ is the volume form on
$CP^m$, where by $\omega^m$ we mean the wedge product taken $m$ times
\cite{law}.  For maps $\Phi:R^{2m} \to CP^m$ there is a topological
winding number $Q_m$, whose expression in local coordinates is
\be
Q_m = {1\over2^m m!}\int_{R^{2m}}d^{2m}x g_{ab}
\({\d\Phibar_a\over\d x^{\mu_1}} {\d\Phibar_b\over\d x^{\nu_1}}
\cdots {\d\Phibar_a\over\d x^{\mu_m}}
{\d\Phibar_b\over\d x^{\nu_m}}\)\epsilon^{\mu_1\nu_1\ldots\mu_m\nu_m}
\ee
This winding number is an integer for fields tending to a constant at
spatial infinity in $R^{2m}$, and it measures the number of times
$\Phi(x)$ covers the target $CP^m$.  These integrals are just the Chern
classes \cite{FibBun}.  $Q_2$ is the 4D abelian anomaly, while $Q_1$
is the total flux in the plane, in units of $2\pi/e$.  Thus for the
$CP^1$ \sm, the flux through any 2-surface $\Sigma$ measures the area
of the target space covered by the image of $\Phi$.  In higher
dimensional target spaces the relation between the restriction of the
K\"ahler form to $\Phi(\Sigma)$ and the induced volume form $dV$ breaks
down.  This is a consequence of Wirtinger's Inequality, which states
that for a 2k-dimensional submanifold $S_k$ of $M$, $\omega^k/k! \le
dV_k$, \cite{law}.  The equality holds only if $\Phi(S_k)$ is a
complex submanifold of the target space, which is a severe restriction.
For 2D submanifolds it amounts to demanding that $\Phi_a$ can be
written as $f(x)(1,w_1,\ldots,w_m)$, with the $w_\alpha$ holomorphic
(or antiholomorphic) functions of a complex coordinate $z=x+iy$ on
$\Sigma$.

Our effective lagrangian (\ref{e:efflag}) reveals that although the
manifold of minima $M$ has the topology of $S^{2N-1}$, as far as the
global degrees of freedom are concerned it is really $CP^{N-1}$.  This
is the result of the existence of a special set of circles in $M$, those
generated by gauge transformations.  Every point in $M$ lies on such a
circle, and in fact this set of circles forms a fibration of
$S^{2N-1}$, the famous Hopf fibration \cite{FibBun}.  The
canonical example occurs at $N=2$, for $CP^1 \simeq S^2$, and it is
well known that $S^3$ can be constructed as a twisted bundle with fibre
$S^1$ and base space $S^2$ (or vice versa, of course).  The physical
significance of these fibres is that the scalar field can lie along
them without costing any gradient energy, for any derivatives can be
counterbalanced by a gauge potential.  This is the underlying reason
for the existence of semilocal strings, for the condition of finite
energy per unit length forces the field to lie on a Hopf fibre at
infinity.  The number of times $n$ this fibre is traversed is given by
\be
n = -{i\over 4\pi} \int_{S^1_\infty}dx^i
{\Phibar{\dbw}_{\! i}\Phi\over|\Phi|^2} =
{e\over 2\pi}\int_{S^1_\infty}dx^iA_i
\ee
The second equality follows from the fact that $a_i$ must vanish at
infinity for finite energy solutions.  The last term is of course
nothing but the total magnetic flux through the plane in units of
$2\pi/e$, which we saw previously was equal to the number of times the
image of $\Phi$ overs the target $CP^m$.  We now see explicitly that
these winding numbers are one and the same,  even away from the low
momentum limit.  The integral of the magnetic flux is $\half\int
d^2x({e}^{-1}\omega_{ij}+f_{ij})\epsilon_{ij}$. For finite energy
solutions the second term must vanish because of the boundary condition
imposed on $a_i$ at infinity.  This leaves an expression which may be
integrated by parts to reveal $2\pi n/e$.  One may legitimately object
that $\omega_{ij}$ is not defined at points where $\Phi=0$.  However,
it is defined for every neighbouring field configuration in which
$\Phi$ vanishes nowhere, and for these configurations the equality
holds between the fibre winding number and the degree of the map from
$R^2$ to $CP^m$. Thus we can consistently extend the definition of
$Q_1$ to maps $\Phi$ which vanish somewhere.

One of the principal reasons for the interest in $CP^m$ \sm s is that
they possess instanton solutions in 2D, which have an arbitrary size
due to the conformal invariance of the theory.  The analogy with 4D
Yang-Mills theories, coupled with the nice properties of 2 dimensions,
make these \sm s ideal laboratories for instanton physics.  In fact, as
we shall see, these instantons are just semilocal strings in the large
scale limit.  The $CP^m$ instanton solutions can be written down very
simply in closed form.  A single instanton is \cite{raj}
\be
\Phi = {(z-z_0)\Phi_1 + w\Phi_2\over\[|z-z_0|^2 + |w|^2\]^{1/2}}
\ee
where $z_0$ and $w$ are arbitrary complex parameters and
$\Phibar_1\Phi_2 = 0$.  This solution represents an instanton centred
at $z_0$ with size $|w|$.  The similarity to the EAH model vortices in
the \bogo\  limit is especially clear if we expand the instanton field
at large distances and compare with (\ref{e:SLasymp}):
\be
\Phi =
\Phi_1{z\over|z|}\(1-\half{|w|^2\over|z|^2}+\frac{1}{4}{|w|^4\over|z|^4}\)
+ \Phi_2 {w\over |z|}\(1-\half{|w|^2\over|z|^2}\)
\ee
Note that at infinity the field tends to $\Phi_1 e^{i\theta}$, which is
a single point in $CP^m$, and identical to the vortex configuration.
The differences only start to appear at 4th degree  in $|w|/|z|$ in this
expansion.  Perhaps the biggest difference is that vortices exist in
the EAH model even at $w=0$, where the $CP^m$ lump is singular.  This
singularity causes problems in numerical simulations of the scattering
of $CP^m$ lumps \cite{lees}, and so the low energy scattering of semilocal
vortices in 2D should be far better behaved \cite{sam-lees}.

To conclude this section we specialise to the case $N=2$, where the
vacuum manifold is $S^3$ and the \sm\  target space is $CP^1 \simeq
S^2$.  This isomorphism is realised through the Pauli matrices
$\sigma^A$, which enables us to construct three real fields
\be
\phi^A = \Phibar\sigma^A\Phi/\eta
\ee
The effective lagrangian (\ref{e:efflag}) can be rewritten in terms on
these scalar fields, and is
\be
{\cal L}_{O(3)} = \frac{1}{4}\d_\mu\phi^A\d^\mu\phi^A -
{1\over 128e^2\eta^2} \[(\d_\mu\phi^A)^2(\d_\nu\phi^B)^2 -
(\d_\mu\phi^A\d^\mu\phi^B)^2\]
\label{E:efflago3}
\ee
This may be verified using the relations
$\sigma^A_{\dot{a}a}\sigma^A_{\dot{b}b} =
2\delta_{\dot{a}b}\delta_{\dot{b}a} - \delta_{\dot{a}a}\delta_{\dot{b}b}$,
and imposing the constraint $|\Phi|^2 = \eta^2$.  Equation
(\ref{E:efflago3}) is the lagrangian for the $O(3)$ \sm\  with a Skyrme
term.  We shall see in the next section that the equivalence between
the \sm s is only local and can break down in an interesting way in
more than 2 dimensions.

\section{Monopoles in Extended Abelian Higgs Models}
The fact that $\pi_2(CP^m)=Z$ naturally leads to the question of
whether monopoles exist in the theory.  Since the coordinates on $CP^m$
are ungauged degrees of freedom, these would have to be a type of
global monopole.  The properties of the EAH model monopoles are best
illustrated in the case $m=1$ through the equivalence to the $O(3)$
\sm\  \cite{raj}, which is well known to possess singular global
monopole solutions  \cite{GlobMono}.  If, however, we view the
constraint $\phi^A\phi^A = \eta^2$ as being imposed by a potential,
{\em i.e.}
\be
{\cal L}_{O(3)} = \frac{1}{4}\d_\mu\phi^A\d^\mu\phi^A -
\frac{1}{8}\lambda(\phi^A\phi^A-\eta^2)^2
\label{e:o3higgs}
\ee
then the theory possesses well-behaved monopole solutions of the form
\be
\phi^A = \eta\hat{x}^Ah(r)
\label{e:hh}
\ee
where $r^2 = x^ix^i$, $h(0)=0$ and $h(\infty)=1$.  The energy of these
solutions is linearly divergent, because the energy density at large
distances is simply $\eta^2/2r^2$, due to the angular variation of the
field.  (In the gauge monopole \cite{Mono} the angular derivatives
are cancelled by the gauge potential.)  Thus if $R$ is a large radius
cut-off, the energy of a monopole is
\be
E_M = 2\pi R\eta^2 + E_c
\label{e:menergy}
\ee
where $E_c$ is the contribution from the centre of the monopole at
which $|\phi| \to 0$.  The energy divergence makes the discussion of a
single global monopole slightly problematic, and dependent on boundary
conditions.  For example, if the centre of the monopole is kept fixed,
there is a cylindrically symmetric zero mode which takes the monopole
away from the hedgehog configuration of (\ref{e:hh}) \cite{gol}.
There is in fact a 1-parameter family of azimuthally symmetric monopole
field configurations, all with the same energy (\ref{e:menergy}).

A physical cut-off would be imposed, for example, by a neighbouring
antimonopole at a distance $\null\sim R$.  Since $dE_M/dR$ is
independent of $R$, the force between the pair of poles can be argued
to be generally independent of their separation (although it also
depends on the details of the field between them \cite{per91}), which
makes them very efficient at annihilating in the early universe
\cite{GlobMono}.

What would such a configuration look like in the $CP^1$ coordinates
$\Phi$?  It is easy to check that the field configuration which gives
(\ref{e:hh}) at large distances is
\be
\Phi = \eta\(\begin{array}{c}
                       \sin\half\theta \, e^{i\varphi} \\
                       \cos\half\theta
          \end{array}\)
\label{e:cp1mono}
\ee
where $\{\theta,\phi\}$ are polar
coordinates.  However, this configuration is ill-defined at $\theta =
\pi$.  In $CP^1$ there is no sign of this problem, since
$(e^{i\varphi},0)$ is just a single point.  There are two ways of
resolving the issue.  The first is to treat it as a gauge artifact, and
define $\Phi$ in two different coordinate patches covering the upper
and lower hemispheres of $S^2_\infty$ (the sphere at $r=\infty$) \cite{Mono}.
Equation (\ref{e:cp1mono}) would be the field on the upper hemisphere
$\Phi_U$, while the field on the lower would be
\be
\Phi_L = \(\begin{array}{c}
                       \sin\half\theta \\
                       \cos\half\theta \, e^{-i\varphi}
          \end{array}\)
\ee
It is easy to see that $\Phi_U$ and $\Phi_L$ are related by a
(singular) gauge transformation where they overlap.  The other
resolution of the problem is to look for solutions in which $\Phi$
vanishes at $\theta=\pi$ in order to avoid the singularity.  With this
departure from spherical symmetry the full solution should have the
form
\be
\Phi = \eta\(\begin{array}{c}
                       h_1(r,\theta)\sin\half\theta\, e^{i\varphi} \\
                       h_2(r,\theta)\cos\half\theta
          \end{array}\)
\ee
where $h_1=0=h_2$ at $r=0$ and $h_1(r,\pi)=0$.  As we approach the $z$
axis near $\theta = \pi$ and at large $r$ the fields should be more or
less $z$ independent.  Thus the top component of $\Phi$ should behave
more or less like the field of a \NO\  vortex, as it obeys similar field
equations and changes phase by $2\pi$ around the axis.  The bottom
component vanishes as $\rho^2/r^2$, where $\rho^2=r^2-z^2$.  Thus we
find a string on the $-z$ axis.  To support this interpretation, let us
compute the gauge potential and the magnetic field.  At large
distances, away from the string, $\Phi$ should lie in its vacuum
manifold and its covariant derivative should vanish, which means that
\be
A_i = \hat{\varphi}_i {\sin^2\half\theta\over er\sin\theta}
\label{e:Amon}
\ee
where $\hat{\varphi}_i$ is a unit vector in the azimuthal direction.
Near $\theta=\pi$, $A_\varphi$ blows up as $1/(e\rho)$.  This is
precisely the field that would be produced by a string on the $-z$
axis, with flux $2\pi/e$ in the upward direction.  The magnetic field
resulting from (\ref{e:Amon})  is
\be
B_i = \hat{r}_i {1\over 2er^2}
\ee
This is a rather strange result.  We started by writing down a global
monopole configuration: from it the EAH model has produced a local
monopole, with one Dirac unit of magnetic charge $2\pi/e$ \cite{Mono}.  This
remarkable object is in fact the $\theta_W = \pi/2$ limit of a field
configuration first singled out by Nambu in the Electroweak theory,
which is essentially a string carrying $Z^0$ flux terminating on a
(electro)magnetic monopole \cite{nam}.  He proposed that an approximately
stable classical solution might be obtained by spinning a finite length
of string with opposite poles at each end.

If the string at $-z$ is viewed as a gauge artifact, in other words as
a Dirac string, then we have constructed a truly spherically
symmetric field configuration, albeit at the price of violating one of
Maxwell's equation, $\d_iB_i=0$.  Consistency can be maintained
either by invoking General Relativity to hide the source of flux in a
black hole, as has been investigated by Gibbons et al. \cite{gib+}, or by
embedding the theory in one with a larger non-Abelian symmetry, so that
the source of the flux is a 't Hooft-Polyakov monopole.

In considering the string/monopole system we are again faced with the
problem of infinite energy common to all global monopoles.  The energy
function, cut off at a distance $R$, can be split into three
components:  the monopole core, the string, and the long range monopole
field.  The energy can then be written
\be
E = E_c + 2\pi\eta^2\nu(\beta)R + 2\pi\rho\eta^2R + \pi/e^2R_c
\ee
where $\nu(\beta)$ is a monotonically increasing function, having the
value 1 at $\beta=1$ \cite{NO}.  The third term is the energy in the
scalar field gradients, with $\rho$ a number depending on the exact
field configuration, but which is known to be bounded below by 1
\cite{alm-lie}.  This bound is saturated when all the scalar field
gradients are concentrated on a line starting at the pole.  The fourth
term is the energy in the magnetic field,
with $R_c$ the monopole core radius.  For sufficiently large $R$ the
magnetic field energy is negligible.  We can try to circumvent the
infinite energy problem by considering a monopole-antimonopole pair,
whose strings run along the $z$ axis from $-R/2$ to $-\infty$ and from
$R/2$ to $+\infty$ respectively.  Although this
configuration has infinite energy also, it is meaningful to talk of the
difference in energy $\Delta E$ between it and a straight string on the
axis.  Naively, this energy difference is just twice the energy of a
monopole with cut-off $R$, minus the energy of the missing piece of
string, or
\be
\Delta E \simeq 2E_c + 2\pi\rho\eta^2R - 2\pi\nu(\beta)R
\ee
Thus for Type I vortices ($\beta<1$) the energy is minimised when
$R=0$, whereas for Type II vortices ($\beta>1$) the monopole separation
can increase without bound, provided a suitable configuration is chosen
for the global degrees of freedom.  This behaviour is consistent
with the results concerning the stability of the vortices as a function
of $\beta$ in Section 2.  The creation of a pair of poles in a straight
string is essentially a very large perturbation in the scalar field
away from the \NO\ string in the region $-R/2<z<R/2$.  We saw that at
$\beta<1$ extended non-singular vortex configurations tended to
collapse towards the \NO\  solution, and it is encouraging that this
heuristic analysis indicates in another way the reappearance of the
\NO\  string.  Numerical work \cite{vac+92} also seems to confirm the
tendency of non-singular vortices to collapse or expand according to
the sign of $\beta-1$.

\section{Textures and Skyrmions}
To complete the study of topological defects in EAH models, we consider
two related types of defects that occur when the third homotopy group
of the vacuum manifold, $\pi_3$, is non-trivial.  They are textures
\cite{Tex} and skyrmions \cite{Skyrm}, which are field
configurations which map the whole of 3 dimensional space, with the
point at infinity included, onto a non-contractible 3-sphere in the
vacuum manifold $M$.  A simple example is the $O(4)$ Goldstone model,
whose Lagrangian is essentially (\ref{e:o3higgs}) but with 4 real
fields.  A spherically symmetric field configuration which covers $M$
once is
\be
\begin{array}{ll}
\phi_1 = \cos\chi & \phi_2 = \sin\chi\cos\theta \\
\phi_3 = \sin\chi\sin\theta\cos\varphi & \phi_4 =
\sin\chi\sin\theta\cos\varphi
\end{array}
\label{e:tex}
\ee
where $\chi(r,t)$ vanishes at $r=0$ and reaches $\pi$ at $r=\infty$.
Exact solutions to the \sm\  equations of motion were found by Spergel
and Turok \cite{Tex}, which collapse from infinite size at $t=-\infty$ to a
singular point and then rexpand towards $t=+\infty$.  At the point of
collapse, the gradient energy density becomes sufficiently large to
force the scalar field through $\phi=0$, which removes the topological
winding.  The dynamic nature of this object is a necessary consequence
of Derrick's scaling argument in 3D \cite{der}, which  shows that both the
gradient and potential terms in (\ref{e:o3higgs}) decrease with scale,
and so any localised object can decrease its energy by shrinking.  If,
however, we supplement the energy functional by a term which is fourth
order in derivatives,
\be
E_4 = \int d^3x\[g_1(\d_i\phi^A)^2(\d_j\phi^B)^2 + g_2
(\d_i\phi^A\d_i\phi^B)^2\]
\ee
then the possibility arises of a stable minimum in the energy for
configurations such as (\ref{e:tex}), because $|E_4|$ increases
linearly with decreasing scale.  If $g_1+g_2=0$, the action remains
second order in time derivatives, and $E_4$ is positive.  This favoured
choice is the so-called Skryme term \cite{Skyrm}.  The importance of skyrmions
lies in their providing a low energy model of the nucleon, where the
Goldstone bosons are interpreted as pions and the topological winding
number as baryon number.

The vacuum manifold of an EAH model with $N$ scalars is $S^{2N-1}$, and
so the only one with textures or skyrmions has $N=2$.  (It is well
known that $\pi_k(S^n)=0$ for $k<n$ \cite{FibBun}.)
We have already seen that the
low momentum dynamics of the model are equivalent to the $CP^1$ \sm,
which is in turn equivalent to the $O(3)$ \sm.  Thus we can expect the
large textures of the $N=2$ EAH model to behave in a very similar way
to the textures in nematic liquid crystals, since their effective
theory is a $S^2/Z_2$ \sm\  \cite{LC}.  In particular, we cannot expect
spherical symmetry.

In the \sm\  approximation to the theory, we can parameterise the
constrained scalar field by angles $\zeta$, $\psi$, and $\sigma$ as
follows:
\be
\Phi = \eta \(\begin{array}{c}
                 \sin\half\zeta e^{i(\sigma+\psi)/2} \\
                 \cos\half\zeta e^{i(\sigma-\psi)/2}
              \end{array}\)
\ee
It is easy to show from $\phi^A = \Phibar\sigma^A\Phi/\eta$ that
$\zeta$ and $\psi$ are polar coordinates on the global $CP^1$ target
space.  The gauge potential that results from ensuring that the current
vanishes everywhere is
\be
A_\mu = \frac{1}{2e}(\cos\zeta\d_\mu\psi + \d_\mu\sigma)
\ee
showing explicitly that $\sigma$ is just a gauge mode.  The gauge field
is then
\be
F_{\mu\nu} = \frac{1}{2e}\sin\zeta(\d_{(\mu}\psi\d_{\nu)}\zeta)
\ee
At this point it is instructive to recall the remarks made in section 3
concerning the geometrical interpretation of $F_{\mu\nu}$ as being (up
to a factor of $e$) the volume form on the image of $\Phi$ in the
target $CP^{N-1}$.  In 2D, it was shown that the flux though a surface
$\Sigma$ was proportional to the area of the $\Phi(\Sigma)$.  Here, we
can display the equality very clearly
\be
\int_\Sigma \half\epsilon^{\mu\nu}F_{\mu\nu}d^2 x =
\frac{1}{e}  \int_{\Phi(\Sigma)} \sin\zeta d\zeta d\psi
\ee
The right hand side is easily seen to be the area on the target space
for the global degrees of freedom, $S^2$.

In studying textures and skyrmions in this theory, it is not entirely
clear which coordinate system to use, for one can no longer assume
spherical symmetry for the objects.   Skyrmions in the $O(3)$ \sm\
have been studied by de Vega and were rediscovered later
by Wu and Zee \cite{deV}.  They both exhibited toroidally
symmetric configurations which covered the target $S^2$ exactly once,
namely
\be
\zeta = \zeta(\mu) \qquad \psi = \beta-\varphi
\ee
where $\{\mu,\beta,\phi\}$ are toroidal coordinates \cite{mor-fes}.
The boundary conditions are $\zeta(0) = 0$ and $\zeta(\infty)=\pi$.
However, this cannot be the basis for a toroidally symmetric solution,
because in these coordinates the second order part of the energy
functional is
\be
E_2 = 2\pi R\eta^2\int d\mu d\beta \frac{\sinh\mu}{(\cosh\mu-\cos\beta)}
\[\(\frac{d\zeta}{d\mu}\)^2 + \sin^2\zeta\coth^2\mu\]
\ee
where $R$ is the radius of the circle $\mu=\infty$.  The explicit
dependence on $\beta$ shows the failure of the toroidal ansatz, which
includes the spherically symmetric case through the stereographic
projection $\cos\zeta = 1-2{\rm sech}^2\mu$.

Nonetheless, it is illustrative to compute, in the low momentum limit,
the magnetic field that this configuration produces.  Defining $\tau$
to be $\cosh\mu-\cos\beta$, we find
\begin{eqnarray}
A_i &=& \frac{1}{2eR}\cos\zeta(\hat{\beta}_i - \hat{\varphi}_i) \\
B_i &=& \frac{\tau}{2eR^2}\[\hat\beta_i\frac{\tau}{\sinh\mu}
\frac{d}{d\mu}\(\frac{\sinh\mu}{\tau}\cos\zeta\) + \hat\varphi_i
\tau\frac{d}{d\mu}\(\frac{1}{\tau}\cos\zeta\)\]
\end{eqnarray}
(Some useful formulae to aid this computation can be found in Ref.
\cite{hua-tip}.)
The lines of flux encircle both the $z$ axis and the circle
$\mu=\infty$ \cite{Twist}.  The flux crossing any surface of constant
$\beta$, which is a spherical bowl with $\mu=\infty$ as its border, is
$2\pi/e$, while the flux crossing surfaces of constant $\varphi$, which
are half planes with the $z$ axis as boundary, is also $2\pi/e$.  Thus
this field configuration can be interpreted as a twisted vortex ring.
In fact, the lines of flux trace the Hopf fibres in the projection $S^3
\to R^3$ that this texture embodies.

There are two topological invariants which characterise the textures
and Skyrmions.  There is the Hopf number
\cite{FibBun,wil-zee,wu-zee}, which measures the degree of
the map $\Phi:R^3\cup\,\infty \simeq S^3 \to S^2$, and is given by
\be
N_H = \frac{1}{8\pi^2} \int d^3x \epsilon^{ijk}\Phibar{\dbw}_i \Phi
\d_j\Phibar_a g_{ab}\d_k \Phi_b/|\Phi|^2
\label{e:hopf}
\ee
where $g_{ab}$ is the Fubini-Study metric on $CP^{N-1}$.  There is also the
Chern-Simons number
\be
N_{CS} = \frac{e^2}{2\pi}\int d^3x \epsilon^{ijk} A_i F_{jk}
\ee
In the low momentum limit, where the gauge field is determined entirely
by the scalar field, these two quantities are equal, as equations
(\ref{e:A}) and (\ref{e:F}) quickly show.  For the EAH model texture
$N_H=N_{CS}=1$.

The toroidal ansatz is particularly good for picking out the way the
energy behaves under a change of scale.  The energy in the magnetic
field is
\be
E_4 = \frac{\pi}{4e^2R}\int d\mu d\beta \frac{\tau}{\sinh\mu}
\(\[\frac{d}{d\mu}\(\frac{\sinh\mu}{\tau}\cos\zeta\)\]^2 +
\sinh^2\mu\[\frac{d}{d\mu}\(\frac{\cos\zeta}{\tau}\)\]^2\)
\ee
$E_4$ clearly behaves in the opposite way to $E_2$, showing that the
Skyrme term can stabilise the texture in the \sm\  approximation,
$\beta\to\infty$.  Comparison of the relative sizes of the second and
fourth order terms shows that the skyrmion has size of order $1/e\eta$,
implying a mass of order $\eta/e$.  Away from the limit, it becomes
possible for the scalar field to be forced through $\Phi=0$ if its
gradient energy density gets too large.  It is therefore an open
question whether a skyrmion exists in the $N=2$ EAH model for finite
$\beta$.

A quantum skyrmion will at best be merely metastable, since the field
can tunnel through the barrier which prevents unwinding, namely the
potential $V(\Phi)$.  The Euclidean action for such an event is of
order $V(0)(e\eta)^{-4}$, and so the skyrmion decay rate is estimated
to be
\be
\Gamma_s \sim (e\eta) e^{-\gamma\beta/e^2}
\label{e:St}
\ee
where $\gamma$ is a numerical coefficient.  In view of the possibility
of factors of $2\pi$ appearing from the symmetries of the skyrmion,
$\gamma$ need not be merely of order unity.

If there is no stable skyrmion then the unwinding results from the
classical evolution of the collapsing texture \cite{TexCollapse}.  In
either case, the magnetic field will then dissipate, as there are no
longer any scalar field gradients to support it, and the Chern-Simons
number changes from 1 to 0.  If there are any anomalous fermionic
currents coupled to our model, there will be a violation of charge
conservation associated with the unwinding event.  This process can be
identified with the manufacture and decay of a sphaleron in the
Electroweak theory in the limit  $\theta_W = \pi/2$.

\section{Cosmological Phase Transitions}
We now turn to the possibility of the creation of any or all of the
defects described in the preceding sections.  It was pointed out by
Kibble \cite{kib76} that when the cooling universe goes through a
phase transition the order parameter (which here is $\Phi$) is
uncorrelated beyond some length scale $\xi$.  At a second order
transition, this scale is determined by the correlation length of the
scalar field at the temperature at which thermal fluctuations cease to
explore the vicinity of $\Phi=0$, or the Ginzburg temperature
\cite{gin}.  At a first order transition it is the average separation
of the bubbles which nucleate the transition which sets the scale.
Thus different domains in space are mapped to different regions of the
manifold of minima of the finite temperature effective potential,
$M_T$.  If this manifold has non-trivial homotopy groups $\pi_0$,
$\pi_1$ or $\pi_2$ then it is possible to produce defects (walls,
strings and monopoles respectively) at the boundaries of the correlated
domains, where the continuity of the order parameter forces it to
vanish.

One can estimate the probability of defect formation by triangulating
both space and $M_T$, and then assigning points in $M_T$ at random to
points on the spatial lattice \cite{vac-vil84}.  This simulates the
appearance of domains of size $\xi$ at the phase transition.  When
$M_T$ has simple geometry, such as that of a sphere, it is possible to
calculate the probabilities without triangulation, just by assigning
points on the sphere to the vertices of the spatial lattice and
interpolating geodesically between them \cite{lees-pro}.  This method
seems more accurate as it does not involve a drastic truncation of the
field configuration space.  For example, let us recall the calculation
of the probability per unit correlation volume of forming monopoles,
where the target manifold is $S^2$.  Firstly, we triangulate $M_T$ with
a tetrahedron.  A monopole is formed if the boundary of a 3-simplex in
the spatial triangulation, which is of course a tetrahedron, is mapped
onto the tetrahedron approximating $M_T$.  There is then a topological
obstruction to mapping the interior of the spatial 3-simplex into
$M_T$, which means that the order parameter must leave $M_T$ and vanish
somewhere inside.  This is the monopole.  The total number of different
ways of mapping the vertices of the spatial triangulation into $M_T$ is
$4^4 = 256$.  The number of these mappings which are mappings of one
tetrahedron onto another is just 24, the order of the tetrahedral
group, including reflections for antimonopoles.  By this method,
therefore, the probability of forming a monopole or an antimonopole is
24/256 = 3/32.  In the alternative method \cite{lees-pro}, the image
of the vertices of a spatial 3-simplex forms 4 random points on the
target manifold.  The gradient energy density of the field is minimised
if the vertices are joined by geodesics, and the edges filled in by the
smaller of the two possible spherical triangles.  A monopole is formed
if the sphere is covered by this construction.  To evaluate the
probability of covering the sphere, consider any three of the four
image points, plus their connecting geodesics, extended to great
circles.  These circles divide the sphere into 8 spherical triangles.
The sphere is covered by the boundary of the spatial 3-simplex if (and
only if) the fourth point is contained within the triangle antipodal to
the one defined by the original 3 points.  The probability of this
happening is proportional to the average fractional area of a spherical
triangle.  By this construction the probability is seen to be 1/8.

In the case at hand the situation is rather more complicated.  We wish
to assess the formation probability of three types of object:  strings,
monopoles, and in the $N=2$ case, texture.  For the first two, the
relevant topological quantity is the magnetic flux through a surface
$\Sigma$.  If the surface is closed, the Maxwell
equation $\d_i B_i = 0$ tells us that the total flux through the
surface is zero.  There are no true magnetic monopoles in this theory,
only approximately spherically symmetric objects which are the termini
of strings.  If the surface has a boundary, the total flux $\F$ tells
us how many strings (more strictly, the difference between the numbers
of strings and antistrings) could pass through that surface.
This is just the integer part of $e\F/2\pi$, say $n$.  However, unlike
the ordinary Abelian Higgs model, there is no topological condition
which forces the scalar field to vanish somewhere on this surface,
since its manifold of minima is simply connected.  Nevertheless, when
$\beta<1$, the configuration near the surface $\Sigma$ with least
energy will consist of a set of \NO\ type vortices with total winding
number $n$, plus any stray flux left over.  In a statistical ensemble
like that produced at a phase transition, however, the most likely
configuration is the one with the minimum {\em free} energy, so it is
not necessarily correct to conclude that the flux is everywhere
confined in vortices.  In fact, there is a lot of entropy in short
lengths of string terminating on monopoles, so it seems probable that
the phase transition produces string segments rather than infinite
string.

Let us now compute the flux through a surface $\Sigma$ with dimensions
of the correlation length $\xi$.  As usual, we triangulate space with a
simplicial complex whose average edge length is $\xi$, and choose
$\Sigma$ to be one of the 2-simplices of the complex.  In using the
topological properties of $M_T$, we are assuming  that the gauge field
is entirely determined by the gradients of the scalar field, which
becomes an increasingly good approximation in the limit $e\to\infty$.
In this limit thermal fluctuations cannot generate real gauge bosons as
they are too masssive.  We now assign values of the scalar field,
constrained to lie in $M_T$, at each vertex.  The field is extended to
all points in $R^3$ by joining the images of the vertices in $M_T$ with
the shortest possible geodesic.  Since $M_T$ is simply connected there
is no topological obstruction to then filling in the resulting set of
geodesic line segments with geodesic surfaces, so that $\Phi$ is
defined everywhere on $\Sigma$.  The task of computing the magnetic
flux through this surface is simple when $N=2$, for as we saw in
Section 3 it is proportional to the area of the image of the simplex,
as measured by the Fubini-Study metric on $CP^{1}$.  Now $CP^1$ is
isomorphic to $S^2$, so the average fractional area occupied by the
image of a spatial 2-simplex is 1/8, exactly as for the monopole
calculation.  If the surface were to cover the whole of the target
$CP^1$, as in the vortex, the total flux would be $2\pi/e$.  Hence the
average flux through a 2-simplex is $\pi/4e$.  The flux through
neighbouring 2-simplices is by construction uncorrelated, so that means
we need an average of $8^2=64$ contiguous 2-simplices before collecting
enough magnetic flux to form a vortex.  Then when $\beta<1$ and
vortices are stable, we expect the string density to be suppressed by a
factor of $8^3$ relative to the $N=1$ case \cite{vac-vil84}.

When there are more than 2 components to the scalar field the
calculation is not so straightforward, because the integral of the
magnetic field over a 2-simplex is no longer equal to the area of its
image in $CP^{N-1}$.  As was discussed in Section 3, by Wirtinger's
inequality the flux is almost always less than the area.

Let us now turn to the probability of texture formation.  First we
reprise the calculation for pure global texture on a vacuum manifold
isomorphic to $S^3$.   The simplest triangulation of $M_T$ is into the
boundary of a 4-simplex, a 4D tetrahedron, which has 5 3-simplices, 10
2-simplices, 10 edges and 5 vertices.  What is required is the average
topological charge per correlated volume: in other words, the average
volume in the target $S^3$ per spatial 3-simplex.  This gives the
number of texture collapses that will occur as the universe expands,
for it appears from dynamical simulations \cite{TexCollapse} that a
region with charge greater than 1/2 ({\em i.e.} covering more than half
the 3-sphere) will eventually collapse and unwind.  There can only be
charge on a spatial simplex if it maps onto a 3-simplex in $M_T$.  The
probabilty of this is $5!/5^4 = 24/125$.  The  $5!$ assignments are
divided into two equal sets with topological charge $\pm1/5$, so the
average charge is of course zero, but he average of the magnitude of the
charge is $24/625 \simeq 0.0384$.  The alternative calculation
\cite{lees-pro} gives the average volume (and thus the average
topological charge) of the 3-simplex defined by 4 random points as
$2^{-4} = 0.0625$.

For texture in the $N=2$ EAH model, we have to exercise care in
assigning meaning to the topological charge, because of the gauge
invariance in the model.  The gauge invariant quantity which measures
the volume of the 3-sphere is the Hopf number, defined in equation
(\ref{e:hopf}).  One can check \cite{wu-zee} that the integrand, which
we term the Hopf density ${\cal H}$, really is the volume element on
$S^3$ by writing $\Phi^T = (y_1+iy_2,y_3+iy_4)$.  Imposing the
constraint $|\Phi|^2 = 1$ it is found that
\be
{\cal H} = {2\over y_4} \epsilon_{ijk} \d_i y_1 \d_j y_2 \d_k y_3
\ee
The average topological charge density is now computed, as before, by
assigning random values in $S^3$ to the vertices of a spatial
3-simplex.  As before, it is $1/16$.

The evolution of an EAH model after a phase transition seems to be a
rather complex problem, especially for $N=2$.  For large $\beta$ the
scalar field is forced to stay in $M_T$ and we need not worry about the
formation of strings and monopoles as they are unstable.  The dynamics
are those of a $CP^{N-1}$ \sm, which should be tractable in the
large-$N$ limit.  If $O(N)$ \sm s are anything to go by
\cite{tur-spe91}, it seems reasonable to suppose that a scaling
solution should be established, with the correlation length of $\Phi$
increasing in proportion to the Hubble length $H^{-1}$.  Accompanying
the scalar field there will be a long range magnetic field belonging to
a broken the broken gauge symmetry, a peculiar prediction of this class
of models.  However, the strength of this field is negligible, of order
$H^2/e^2$, and it contributes only a tiny fraction $\sim G^2H^2/e^2$ to
the energy density of the universe.  There is an interesting
complication when $N=2$, for here there are textures, a few of which
should collapse per Hubble volume per expansion time \cite{Tex}.  If
$\beta$ is sufficiently large, a population of skyrmions may result, as
the magnetic field pressure prevents the final unwinding of the
``knot'' of scalar field gradients (see Section 5).   Skyrmions, even
at the GUT scale, may represent a serious cosmological problem, for
their lifetime is $\sim 10^{-39}e^{\gamma\beta/4\pi \alpha}$ s
(\ref{e:St}), and we are perforce discussing the large $\beta$ case.
For example, if $\gamma\beta>O(10)$ then the lifetime becomes longer
than 1s, which brings with it the potential for conflict with standard
Big Bang nucleosynthesis \cite{BigBang}.

At low $\beta$, the stability of the vortex solutions opens up the
possibility that they dominate the evolution of the model.  Let us
assume, as argued above, that the phase transition results in the
formation of finite length segmenta of string terminated by the
monopoles of Section 4.  If these were local monopoles one would expect
the strings to disappear in a very short time \cite{Mono} as the string
tension pulled the monopoles together.  However, there is a competing
effect: the linear potential between monopole-antimonopole pairs, which
can act to join two nearby segments into a single longer one.  This
string can be created by the annihilation of global monopoles.  It is
thought that both global monopoles and strings, separately, can reach a
scaling solution in an expanding universe, so it is conceivable that
the $\beta<1$ EAH model can also scale.  At any time, each Hubble
volume would contain a few lengths of string, terminating on
annihilating global monopoles.  The string would decay in the usual way
by loop production and gravitational radiation.

\section{Discussion and Conclusions}
We have seen that EAH models exhibit a surprisingly rich set of
classical solutions.  They also teach some valuable lessons about
defects in field theories.  When there are spontaneously broken local
and global symmetries, the standard test for defects, the existence of
non-trivial homotopy groups of the vacuum manifold, must be applied
with care.  Let the group of local symmetries of the theory be $G_l$,
which must be a subgroup of the group of all the global symmetries G,
since any local symmetry is also a global one.  The action of the gauge
group on the vacuum manifold divides it into a set of orbits, each
isomorphic to the coset space $G_l/H_l$, where $H_l$ is the unbroken
subgroup of $G_l$.  Let us call this coset space $M_l$.  The action of choosing
a gauge picks out a single point on each orbit, and the set of these
points forms a manifold $M_g$, which is the target manifold of the
Nambu-Goldstone fields.  What we have, of course, are all the elements
of a fibre bundle \cite{FibBun}.  The bundle space is $M$, the fibre is
$M_l$, and when we choose a gauge we project down to the base space $M_g$.
If it were assumed that $M$ is globally a direct product $M_l\times M_g$,
the homotopy groups of
$M$ would then be a direct sum $\pi_k(M_l) + \pi_k(M_g)$, and any
non-trivial topology in $M_l$ and $M_g$ would show up in $\pi_k(M)$.
As we have found, however, this is not  a safe assumption to make in
general.  If we had made it for the $N=2$ EAH model, we would have
concluded that $M \simeq S^1\times S^2$, and that the non-trivial
homotopoy groups were a local $\pi_1$, and a global $\pi_2$ and
$\pi_3$.  This would indicate that this theory had gauge vortices,
global monopoles, and global texture.  The reality is somewhat more
complicated:  the gauge vortices are not necessarily stable;  the
global monopoles have a magnetic charge supplied by a real string
attached to it; and the global texture also has an associated magnetic
field.
This is all a result of the fact that it is possible to assemble a
fibre and a base space to make a smooth manifold in more than one way:
in this particular case, $M$ is the Hopf bundle \cite{FibBun}, with
fibre $S^1$ over a base space $S^2$, which is globally isomorphic to
$S^3$.  In view of this feature I offer a definition of a semilocal
defect:  a defect in a theory whose vacuum manifold is a non-trivial
bundle with fibre $G_l/H_l$.

As a result, one must reconsider the statement sometimes made that
$d$-dimensional defects exist in $n$ space dimensions if
$\pi_{n-d-1}(M)$ is non-trivial.  Suppose we are searching for finite
energy defects.  The finite energy condition means that, at spatial
infinity, the field must lie in its vacuum manifold {\em and} be on a
gauge orbit: that is, it must lie in $M_l$.  Thus one must really
examine $\pi_{n-d-1}(G_l/H_l)$.  However, not even the non-triviality
of this group guarantees the existence of a stable static solution to
the field equations, for we saw that the stability of the vortex
solution in the EAH models depends on the sign of $\beta-1$.  Thus a
non-trivial homotopy group reveals only that finite energy field
configurations fall into inequivalent classes, and does not generally
give information about the existence of stable solutions within these
classes.

At the start of the paper, I put forward arguments in favour of models
with mixed local and global symmetries, based on a principle of economy
in Grand Unified Theories.  The class of models considered here have
turned out to have much intrinsic interest, but it would be encouraging
to demonstrate that GUTs really can produce semilocal defects.  Serious
model-bulding is beyond the scope of the present work, a candidate
semilocal grand unified model is at last possible.  Let us try to embed
the local symmetry in a plausible grand unified group, which must
therefore be of at least rank 5.  The most obvious group to try is
$SO(10)$, and its simply connected covering, $Spin(10)$.  The
\NO\ string can be embedded in this theory by first breaking it with an
adjoint to $SU(5)\times U(1)$.  This $U(1)$ may now be broken with a
{\bf 126} dimensional $\phi_{126}$ down to $Z_2$, resulting in stable
strings \cite{kib-laz-sha}.  It is of course possible to break directly
to $SU(5)\times Z_2$ (strictly, $SU(5)\times Z_{10}/Z_5$
\cite{oli-tur}) with the {\bf 126} alone, but the resulting string is
not simply an embedding of the \NO\ vortex \cite{NonAbString}.  The
{\bf 126} contains an $SU(5)$ singlet $\Phi$ which accomplishes the
required symmetry breaking, and it is this component that makes the
string.  In order to give a global symmetry to the string field we must
therefore replicate the $\Phi_{126}$.

Finally, the reader will have noticed that no attempt has been made to
calculate the cosmological perturbations produced by these models.
This is because the calculation seems rather messy (except perhaps in
the limit of large $N$ and large $\beta$ where the \sm\ approximation
works well).  There exist calculations in various stages of refinement
for strings \cite{CMB-String} and texture (Spergel and Turok
\cite{Tex}) separately, but it is not clear how they are combined with
the global monopole signal in EAH models.  It is however worth pointing
out that existing texture calculations of CMB fluctuations assume a
vacuum manifold isomorphic to $S^3$.  A collapsing texture in an $N=2$
EAH model will not give the same distinctive signal, a $10^\circ$ hot
or cold spot on the microwave sky, if only because of the lack of
spherical symmetry.  This may be of importance to the texture galaxy
formation scenario \cite{Tex}, which may be in danger of conflicting with
the recent COBE data \cite{COBE}.

I am grateful to the Institute of Theoretical Physics, Santa Barbara
for hospitality while this work was completed.  I thank Gary Gibbons,
Robert Leese, Ian Moss, Hugh Osborn and Trevor Samols for useful
discussions, and I am particularly indebted to Rich Holman and
Tom Kephart for reading the manuscript, correcting mistakes, and suggesting
improvements.   This work was supported in part by the National Science
Foundation under grant number PHY89-04035.

\nonumber\section{Tables}
TABLE 1: Table of eigenvalues $\epsilon$ of equation (\ref{e:pert}),
for various values of the parameter $\beta$ (see text).\par\nobreak
$$
\vbox{\offinterlineskip
\hrule
\halign{&\vrule#&\strut\quad\hfil#\hfil\quad\cr
height2pt&\omit&&\omit&\cr
&$\beta$&&$\epsilon$&\cr
height2pt&\omit&&\omit&\cr
\noalign{\hrule}
height2pt&\omit&&\omit&\cr
&100&&$-99.234$&\cr
&30&&$-29.208$&\cr
&10&&$-9.165$&\cr
&3&&$-2.088$&\cr
&1&&$3.81\times10^{-3}$&\cr
height2pt&\omit&&\omit&\cr}
\hrule}
$$


\begin{thebibliography}{99}
\bibitem{Twist} A diagram which can be interpreted as a representation
 of the flux lines can be found on the cover
of any edition of the Twistor Newsletter, produced by the Mathematical
Institute, Oxford.
\end{thebibliography}
\end{document}